\documentclass[aps,pre,twocolumn,groupedaddress,superscriptaddress, showpacs]{revtex4-1}

\usepackage{graphicx}

\usepackage[usenames,dvipsnames]{color}

\usepackage{transparent}

\usepackage{tipa}

\usepackage{amsmath}

\usepackage{amsfonts}

\usepackage{amssymb}

\usepackage{dcolumn}

\usepackage[english]{babel}

\usepackage{array}

\usepackage{dsfont}

\begin{document}

\title{Prediction and control of slip-free rotation states in sphere assemblies}

\author{D. V. St\"ager}

  \email{staegerd@ethz.ch}

  \affiliation{Computational Physics for Engineering Materials, IfB, ETH Zurich, Wolfgang-Pauli-Strasse 27, CH-8093 Zurich, Switzerland}

\author{N. A. M. Ara\'ujo}

  \email{nmaraujo@fc.ul.pt}

  \affiliation{Departamento de F\'{\i}sica, Faculdade de Ci\^{e}ncias, Universidade de Lisboa, P-1749-016 Lisboa, Portugal, 
and Centro de F\'isica Te\'orica e Computacional, Universidade de Lisboa, P-1749-016 Lisboa, Portugal}

\author{H. J. Herrmann}

  \email{hans@ifb.baug.ethz.ch}

  \affiliation{Computational Physics for Engineering Materials, IfB, ETH Zurich, Wolfgang-Pauli-Strasse 27, CH-8093 Zurich, Switzerland}
  
  \affiliation{Departamento de F\'isica, Universidade Federal do Cear\'a, 60451-970 Fortaleza, Cear\'a, Brazil}

\begin{abstract}
We study fixed assemblies of touching spheres that can individually rotate. From any initial state, sliding friction drives an assembly toward a slip-free rotation state. For bipartite assemblies, which have only even loops, this state has at least four degrees of freedom. For exactly four degrees of freedom, we analytically predict the final state, which we prove to be independent of the strength of sliding friction, from an arbitrary initial one. With a tabletop experiment, we show how to impose any slip-free rotation state by only controlling two spheres, regardless of the total number.
\end{abstract}

\pacs{05.45.Xt, 89.75.-k, 45.40.Bb, 45.20.dc}

\maketitle

In a fixed assembly of touching spheres that can rotate individually, sliding friction at the contacts between spheres will generally slow down and finally stop the rotation of all spheres. However, if the assembly only contains even loops (bipartite assembly), it will instead drive toward a stationary slip-free rotation state which is free of sliding friction. This state happens to have at least four degrees of freedom. Here we introduce time invariant quantities and with them show, for assemblies with exactly four degrees of freedom, that no matter how many spheres belong to the assembly, one can always predict from the initial state of rotation the final state, which we prove to be independent of the type and strength of sliding friction. This allows imposing any slip-free state by only controlling two spheres, providing a method to control the collective state of rotation of an assembly. With this work, we contribute to the understanding of the rotational dynamics of dense packings which are studied in the context of shear bands \cite{Astrom2008,Astroma2012,Halsey2009,Rivier2011,Astrom2000} and seismic gaps \cite{Astrom2008,Astroma2012,Roux1993,Astrom2000,Herrmann1990,Baram2004,Baram2005}. Since we propose a new way of controlling the rotation of spheres, mechanics and robotics is another area of applicability of our results.

\begin{figure}[h]
\begin{center}
	\includegraphics[width=\columnwidth]{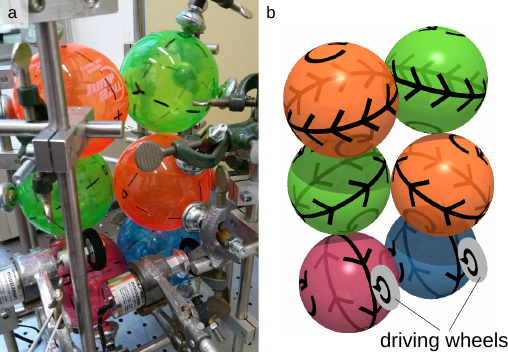}
\end{center}
\caption{
\label{fig:three_pairs} \textbf{Operation of an assembly of spheres in a slip-free state.} (a) Three pairs of plastic spheres in contact were stacked onto each other with an alternating stacking angle of 35 degrees and their positions were fixed using ball rollers. When forcing the bottom pair of spheres to rotate according to the scheme shown in (b) using external driving wheels, the other spheres adjust their rotation due to friction. In the stationary state the top pair of spheres rotates more than three times faster than the bottom pair.
}
\end{figure}

Consider three pairs of plastic spheres in contact alternatively displaced (at a stacking angle) on top of each other, with their positions fixed by ball rollers as shown in Fig.\ \ref{fig:three_pairs}(a). Forcing the two bottom spheres to rotate as indicated in Fig.\ \ref{fig:three_pairs}(b), we find a stationary state of rotation in which the top pair of spheres rotates more than three times faster than the bottom pair (video in the Supplemental Material). Since any two touching spheres have equal tangential velocities at their contact point, this state of rotation is slip-free and, therefore, free of sliding friction. This assembly is bipartite and as we show later, its slip-free state has four degrees of freedom, such that we call it a 4DOF assembly. Here, we first explore the kinetics of bipartite assemblies in general. We then show how to construct 4DOF assemblies, how to predict their final slip-free state from an arbitrary initial one, and how to control the rotation state experimentally. The discovery of the ability to control the rotation state of an assembly was unexpected. Our work is motivated by previous studies on bipartite assemblies of disks \cite{Astrom2008,Halsey2009,Rivier2011,Astroma2012,Manna1991a,Doye2005,
Kranz2015,Roux1993,Herrmann1990,Araujo2013,Astrom2000} and spheres \cite{Baram2004,Baram2005}.

\begin{figure}[t]
\begin{center}
	\includegraphics[width=\columnwidth]{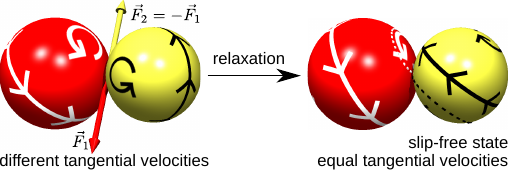}
\end{center}
\caption{
\label{fig:Single_Pair} \textbf{Relaxation of a single pair of spheres toward the slip-free state.} Rotating spheres with sliding forces at their contact (left) and in the final slip-free state (right), in which the contact moves along circles (dashed). Due to Newton's third law of motion \mbox{$\vec{F}_2=-\vec{F}_1$}, where $\vec{F}_1$ and $\vec{F}_2$ are the sliding forces acting on the first and second sphere, respectively.
}
\end{figure}

Let us first consider two single spheres in contact that relax from an arbitrary initial rotation state toward a slip-free state, as shown in Fig.\ \ref{fig:Single_Pair}. In the following, we color any sphere either red or yellow such that no spheres of the same color touch, what is only possible in bipartite assemblies. We assume spheres are perfectly rigid and we only consider sliding friction. As long as the slip-free state is not reached yet, the tangential velocities of the spheres at the contact differ and the two sliding forces at the contact (one on each sphere) tend to reduce this velocity difference. The two forces are opposite to each other and each force $\vec{F}$ produces a torque \mbox{$\vec{T}=\vec{r} \times \vec{F}$}, where $\vec{r}$ points from the center of the corresponding sphere to the contact. The two torques are parallel and their magnitude is proportional to the radius of the corresponding sphere. Using the law of motion \mbox{$\vec{T}= I \vec{\alpha}$}, we find independently of the type and strength of sliding friction
\begin{equation}\label{eq:IwdivR}
\frac{I_1 \vec{\alpha}_1}{r_1}=\frac{I_2 \vec{\alpha}_2}{r_2},
\end{equation}
where $I_1$, $I_2$, $\vec{\alpha}_1$, $\vec{\alpha}_2$, $r_1$, and $r_2$ are the moments of inertia, the angular accelerations, and the radii of the first and the second sphere, respectively. Note that since we assume that the centers of the spheres are fixed, the two sliding forces merely lead to torques which do not cancel each other such that angular momentum is not conserved.

In bipartite assemblies with many spheres, a single sphere might have multiple contacts and multiple simultaneously acting sliding forces. 
Each sliding force contributes to the angular acceleration of the sphere. We define $\vec{\alpha}_i^k$ as the contribution to the angular acceleration $\vec{\alpha}_i$ due to the sliding force at contact $k$, such that \mbox{$\vec{\alpha}_i = \sum_k \vec{\alpha}_i^k$}, the sum running over all contacts. We write Eq.\ (\ref{eq:IwdivR}) analogously for a contact $k$ between spheres $i$ and $j$ with the contributions $\vec{\alpha}_i^k$ and $\vec{\alpha}_j^k$. For simplicity, we consider the moment of inertia of a sphere $i$ proportional to its mass $m_i$ and its radius squared, as, for example, for homogeneous solid spheres. We find 
\begin{equation}\label{eq:mraik-mrajk}
m_i r_i \vec{\alpha}_i^k - m_j r_j \vec{\alpha}_j^k = \vec{0}.
\end{equation}
Summing this equation over all contacts $k$, we obtain
\begin{equation}\label{eq:smRa=0}
\sum_i s_i m_i r_i \vec{\alpha}_i= \vec{0},
\end{equation}
where the sum runs over all spheres and $s_i$ is $+1$ if sphere $i$ is red and $-1$ if it is yellow.
Using Eq.\ (\ref{eq:smRa=0}) we define
 \begin{equation}\label{eq:BearingMomentum}
 \vec{A} := \sum_i s_i m_i r_i \vec{\omega}_i,
 \end{equation}
where $\vec{\omega}_i$ is the angular velocity of the sphere $i$. Equation (\ref{eq:smRa=0}) shows that \mbox{$\partial\vec{A}/\partial t  = \vec{0}$} , i.e., $\vec{A}$ is time invariant. 

We derive one further time invariant quantity. Let $\vec{x}_i$ be the position vector of the center of sphere $i$, where we choose the center of mass $\vec{M}$ of the entire assembly to be the origin of our coordinate system, i.e., \mbox{$\vec{M}=\vec{0}$}. We multiply (dot product) Eq.\ (\ref{eq:mraik-mrajk}) by $\vec{x}_i$ and obtain \mbox{$m_i r_i \vec{\alpha}_i^k \cdot \vec{x}_i - m_j r_j \vec{\alpha}_j^k \cdot \vec{x}_i = 0$}. Since the vector \mbox{$\vec{x}_{ij} = \vec{x}_j - \vec{x}_i$} is perpendicular to $\vec{\alpha}_j^k$, \mbox{$m_j r_j \vec{\alpha}_j^k \cdot \vec{x}_{ij}$} vanishes. We subtract it from the previous equation and get \mbox{$m_i r_i \vec{\alpha}_i^k \cdot \vec{x}_i - m_j r_j \vec{\alpha}_j^k \cdot \vec{x}_j = 0$}. Analogously to the derivation of $\vec{A}$, we thereby define the time invariant quantity $B$ as
\begin{equation}\label{eq:HedgehogQuota}
B:= \sum_i s_i m_i r_i\vec{\omega}_i \cdot \vec{x}_{i}.
\end{equation}

\begin{figure}[t]
\begin{center}
	\includegraphics[width=\columnwidth]{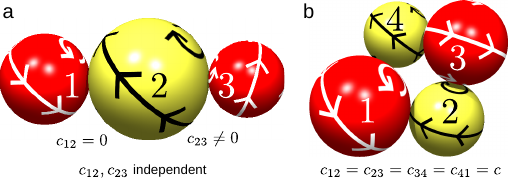}
\end{center}
\caption{
\label{fig:pair_chain_loop} \textbf{Parameters describing the slip-free state.} The $c_{ij}$ parameters that uniquely relate the angular velocities of contacting spheres $i$ and $j$ in the slip-free state are independent for an open chain (a) and are all equal to one single $c$ for a \textit{noncoplanar-4} loop (b), which is a 4DOF assembly. For $c_{ij}=0$, the angular velocities of spheres $i$ and $j$ are antiparallel.
}
\end{figure}

Later we use the quantities $\vec{A} \in \mathds{R}^3$ and $B \in \mathds{R}$ to predict the slip-free state of 4DOF assemblies. Now we identify how 4DOF assemblies can be constructed. The condition for the slip-free state of two spheres $i$ and $j$ is that their tangential velocities at their contact are equal. We formulate this as \mbox{$\vec{\omega}_i^\text{sf} \times r_i \hat{x}_{ij} = \vec{\omega}_j^\text{sf} \times r_j \hat{x}_{ji}$}, where $\vec{\omega}_i^\text{sf}$ is the angular velocity of sphere $i$ in the slip-free state (sf), and $\hat{x}_{ij}$ is the unit vector pointing from sphere $i$ to sphere $j$. We rewrite this condition in analogy to Ref.\ \cite{Baram2004} as
\begin{equation}\label{eq:cij}
s_j r_j \vec{\omega}_j^\text{sf} - s_i r_i \vec{\omega}_i^\text{sf} = c_{ij} \vec{x}_{ij},
\end{equation}
which defines $c_{ij} \in \mathds{R}$ that uniquely relates $\vec{\omega}_i^\text{sf}$ to $\vec{\omega}_j^\text{sf}$. The slip-free state is uniquely defined by the angular velocity $\vec{\omega}_i^\text{sf}$ of a single sphere $i$ and the parameters $c_{ij}$ of all contacting spheres $i$ and $j$. Since $\vec{\omega}_i^\text{sf} \in \mathds{R}^3$, the number of DOFs of the slip-free state is equal to three plus the number of DOFs of the set of $c_{ij}$'s, such that we only have a 4DOF assembly, if the set of $c_{ij}$'s is restricted to a single DOF. Therefore a single pair of touching spheres with a single $c_{ij}$ is the simplest 4DOF assembly. For any longer open chain of spheres, we have an independent $c_{ij}$ for each contact as shown in Fig.\ \ref{fig:pair_chain_loop}(a), not resulting in a 4DOF assembly. 
For a bipartite loop with an even number of spheres $N\geq4$, the $c_{ij}$'s of the contacts are not independent and have $N-2$ DOFs in case all the centers of the spheres are coplanar, and $N-3$ otherwise (details in appendix \ref{appendix:DOFScijs}). Therefore, the only loop that is a 4DOF assembly is a loop of four spheres whose centers are not coplanar (see e.g.\ Fig.\ \ref{fig:pair_chain_loop}(b)), which we denote as a \textit{noncoplanar-4} loop, where all $c_{ij}$'s have to be equal to a single parameter $c$. Starting with a pair of spheres, every other 4DOF assembly can be constructed by iterative extension in the following ways, as long as the assembly is bipartite. Either connect two 4DOF assemblies by establishing two contacts involving two spheres of each assembly such that the centers of the four spheres involved are not coplanar. Or connect a single sphere via two contacts to a 4DOF assembly such that the centers of the involved spheres are not collinear. With these options as minimal requirements, one can even form the space-filling assembly presented in Ref.\ \cite{Baram2004}. Both ways of extension ensure that the $c_{ij}$'s of all contacts are equal to a single $c$ (details in the appendix \ref{appendix:Ensurecijs} and \ref{appendix:DescriptionState}), such that given Eq.\ (\ref{eq:cij}), the slip-free states of 4DOF assemblies can be uniquely described by
\begin{equation}\label{eq:sync_ang_velocities_baram}
\vec{\omega}_i^\text{sf} = \frac{s_i}{r_i} \left( \vec{\Omega} + c \vec{x}_{i}\right),
\end{equation}
where $\vec{\Omega} \in \mathds{R}^3$ is a vectorial reference quantity. 

We can now predict the final state toward a 4DOF assembly will drive by using Eqs.\ (\ref{eq:BearingMomentum}), (\ref{eq:HedgehogQuota}), and (\ref{eq:sync_ang_velocities_baram}) (details in the appendix \ref{appendix:PredictionState}) to find \begin{equation}\label{eq:sync_ang_velocities}
\vec{\omega}_i^\text{sf} = \frac{s_i}{r_i} \left( \frac{\vec{A}}{M} + \frac{B}{I} \vec{x}_{i}\right),
\end{equation}
with $M=\sum_i m_i$ and $I=\sum_i m_i |\vec{x}_i|^2$. Equation (\ref{eq:sync_ang_velocities}) shows that the final slip-free state can be predicted using the time invariant quantities $\vec{A}$ and $B$ defined in Eqs.\ (\ref{eq:BearingMomentum}) and (\ref{eq:HedgehogQuota}). Surprisingly, the final state is independent of the type and strength of sliding friction. Different sliding forces merely lead to different kinetic pathways toward the slip-free state as shown in the appendix \ref{appendix:Differentpathways}. 

\begin{figure}[t!]
\begin{center}
	\includegraphics[width=\columnwidth]{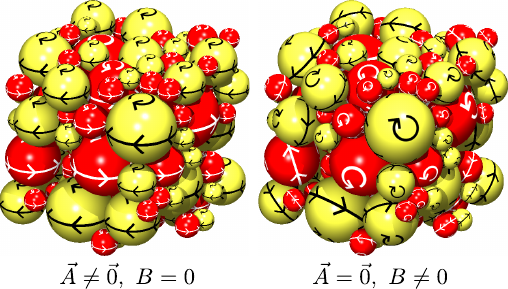}
\end{center}
\caption{
\label{fig:B_H} \textbf{Description of different slip-free states.} 4DOF assembly in different slip-free states. For \mbox{$\vec{A} \neq \vec{0}$} and \mbox{$B = 0$} all axes of rotation are parallel to $\vec{A}$. For \mbox{$B \neq 0$} all axes of rotation of the spheres meet at the position \mbox{$\vec{x}=I \vec{A} / (B M)$}, and in particular for \mbox{$\vec{A}=\vec{0}$} they meet at the center of mass.}
\end{figure}

Figure \ref{fig:B_H} pictures the role of $\vec{A}$ and $B$ regarding the slip-free state predicted by Eq.\ (\ref{eq:sync_ang_velocities}). For \mbox{$B=0$} all axes of rotation are parallel to $\vec{A}$. For \mbox{$B \neq 0$} they all intersect at \mbox{$\vec{x}=I \vec{A} / (B M)$}, and a sphere located at $\vec{x}$ would be at rest and would rotate faster the larger its distance to $\vec{x}$ is.

Remarkably, if one blocks one sphere, generally not all spheres will stop rotating but the assembly will instead relax toward a slip-free state where all axes of rotation intersect at the center of the blocked sphere (details in the appendix \ref{appendix:BlockingAsphere}). Blocking two or more spheres at the same time will force every sphere to stop. Assemblies with more than four DOFs allow more than one sphere to be blocked without stopping all spheres.

To impose any desired slip-free state of a 4DOF assembly within our model, one can apply instantaneous changes to two arbitrarily chosen angular velocities as we show in detail in the appendix \ref{appendix:ImposingState}, where we additionally show how to determine the total mass $M$, the center of mass, and the parameter $I$ by accessing only two spheres (appendix \ref{appendix:DetermineQuantities}). A real system that is subject to e.g., rolling, torsion, and air friction would always come to rest from any initial state. Nevertheless, one can preserve any desired slip-free state of Eq.\ (\ref{eq:sync_ang_velocities}) (compare Fig.\ \ref{fig:B_H}) by preserving two angular velocities accordingly. How close the real stationary state will reach the desired state depends on various details of the assembly such as spatial arrangement, material, size, contact forces between spheres, and the fixing structure. We conducted a simple experiment as a first demonstration on how the reached stationary state might depend on the spatial arrangement.
\begin{figure}[t]
\begin{center}
	\includegraphics[width=\columnwidth]{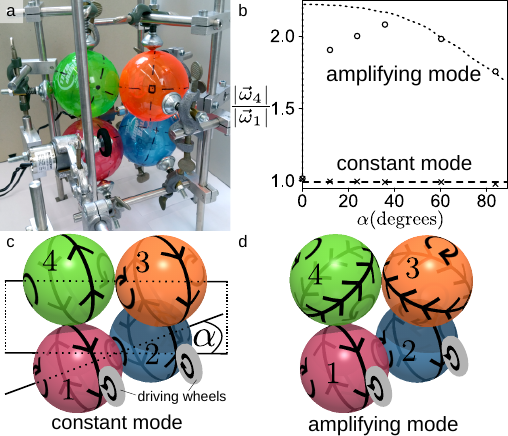}
\end{center}
\caption{
\label{fig:exp} \textbf{Tabletop experiment to compare with the model.} (a) Two pairs of contacting TPU spheres were stacked onto each other and their positions were fixed using ball rollers. The two lower spheres were forced to rotate with equal absolute angular velocity using driving wheels attached to motors. (b) Ratio of the angular velocities of an upper and a lower sphere (experiment (symbols) in comparison with theory (lines)) versus different stacking angles $\alpha$ (compare (c)), for two different modes of rotation indicated in (c) (\emph{constant mode}) and (d) (\emph{amplifying mode}).
}
\end{figure}
Two horizontally contacting pairs of hollow TPU spheres were stacked onto each other as shown in Fig.\ \ref{fig:exp}(a) with varying stacking angle $\alpha$ (compare Fig.\ \ref{fig:exp}(c)), forming a \emph{noncoplanar-4} loop (4DOF assembly) for $0^{\circ}<\alpha<90^{\circ}$, and a \emph{coplanar-4} loop (5DOF assembly) for $\alpha=0^{\circ}$. We fixed the positions of the spheres using ball rollers, allowing the spheres to rotate. The two lower spheres were forced to rotate with equal absolute angular velocity around the axis connecting their centers. When we force them to rotate in opposite directions as shown in Fig.\ \ref{fig:exp}(c), 
we impose a state where all rotation axes are parallel and the absolute angular velocities are equal, such that we call this the \emph{constant mode}. When forcing both to rotate in the same direction as shown in Fig.\ \ref{fig:exp}(d), we impose for $\alpha>0^{\circ}$ a state where all axes of rotation meet at the contact between the two lower spheres. Then the two upper spheres will rotate at a higher angular velocity as we have shown in Fig.\ \ref{fig:three_pairs}; therefore, we call it the \emph{amplifying mode}. 
Figure\ \ref{fig:exp}(b) shows the measured ratio of the angular velocities of an upper and a lower sphere (symbols) as well as the prediction by our model (lines). In the \emph{constant mode} the experiment agrees well with our prediction for all angles $\alpha$. In the \emph{amplifying mode}, we find excellent agreement for $\alpha \approx \{60^{\circ},84^{\circ}\}$. For $\alpha \approx \{12^{\circ},24^{\circ},36^{\circ}\}$, the angular velocities of the upper spheres were on average smaller than the predicted ones, and we did not observe smooth but partially jerky rotations becoming jerkier for decreasing $\alpha$. In the \emph{amplifying mode}, the spatial arrangement for low $\alpha$ triggers this deviation which originates from the elasticity of the spheres. In our experimental setup, elastic deformation can lead to jerky motions by causing displacements of spheres and by the formation of non-negligible contact areas between spheres that can lead to torsional forces as we explain in detail in the appendix \ref{appendix:elasticity}. Experimental details can be found in the appendix \ref{appendix:experimentaldetails} and videos in the Supplemental Material.

This Letter advances the understanding of the rotational degrees of freedom of assemblies and thus of the internal dynamics of shear deformation and seismic gaps. In a technological perspective, the ability to control the rotation state of an assembly of rotating spheres in contact is a newly discovered functionality with a general but yet unexplored potential. It is likely to find use in mechanics and robotics to control the orientation and rotation of spheres. The possibility to amplify the angular velocities of spheres along a certain contact network could be employed as an alternative to power transmission gears. Since the ability to control the rotation state is robust against changes in the spatial arrangement, as long as contacts are conserved, assemblies allow for desired or inevitable displacements during operation.

\begin{acknowledgments}
We acknowledge financial support from the ETH Risk Center, the Brazilian institute INCT-SC, Grant No.\ FP7-319968-FlowCCS of the European Research Council (ERC) Advanced Grant, and the Portuguese Foundation for Science and Technology (FCT) under Contracts
No.\ EXCL/FIS-NAN/0083/2012, No.\ UID/FIS/00618/2013, and No.\ IF/00255/2013. Special thanks goes to Falk Wittel for contributions regarding the implementation and construction of the experiment.
\end{acknowledgments}

\bibliography{bibliography_paper}

\appendix

\clearpage
\newpage

\section{Degrees of freedom of $c_{ij}$'s for loops}\label{appendix:DOFScijs}
Let us consider a bipartite loop with an even number of spheres $N \geq 4$. The angular velocities $\vec{\omega}_i^\text{sf}$ and $\vec{\omega}_j^\text{sf}$ of two contacting spheres $i$ and $j$ in the slip-free state (sf) are uniquely related to each other via a parameter $c_{ij}$ according to
\begin{equation} \label{eqA:c_ij}
s_j r_j \vec{\omega}_j^\text{sf} - s_i r_i \vec{\omega}_i^\text{sf}  = c_{ij} \vec{x}_{ij},
\end{equation}
where $r_i$ is the radius of the sphere $i$, $s_i$ is $+1$ if sphere $i$ is red and $-1$ if it is yellow, and $\vec{x}_{ij}=\vec{x}_{j}-\vec{x}_{i}$ is the vector pointing from the center $\vec{x}_{i}$ of sphere $i$ to the center $\vec{x}_{j}$ of sphere $j$. Since for every contact Eq.\ (\ref{eqA:c_ij}) needs to hold, one finds the constraint

\begin{equation}\label{eqA:constraint}
c_{12} \vec{x}_{12} + c_{23} \vec{x}_{23} + \ldots + c_{N1} \vec{x}_{N1}=\vec{0},
\end{equation}
where the indexes $1$ to $N$ are given to the spheres in consecutive order. The set of $c_{ij}$'s has $N-R$ DOFs, where $R$ is the rank of the $3 \times N$ matrix $\left(\vec{x}_{12} \ \vec{x}_{23} \ \cdots \  \vec{x}_{N1} \right)$. $R$ is equal to the number of linear independent columns of the matrix and is either two, when all centers of the spheres are coplanar, or three otherwise. For \emph{noncoplanar-4} loops, it follows from Eq.\ (\ref{eqA:constraint}) that all $c_{ij}$'s need to be equal and can be replaced by a single parameter $c$.

\section{Ensure that all $c_{ij}$'s are equal}\label{appendix:Ensurecijs}

To extend a 4DOF assembly, one needs to ensure that all $c_{ij}$'s are equal to a single $c$. Starting from any 4DOF assembly, one finds for any pair of contacting spheres $i$ and $j$ that
\begin{equation} \label{eqA:srw_ij}
s_j r_j \vec{\omega}_j^\text{sf} - s_i r_i \vec{\omega}_i^\text{sf}  = c \vec{x}_{ij}.
\end{equation}

Let us first relate the angular velocities of two spheres $i$ and $k$ which are not in contact, but are both touching a third sphere $j$. We sum Eq.\ (\ref{eqA:srw_ij}) and an analogous equation for the contact between sphere $j$ and $k$ to find
\begin{equation}
\begin{aligned}
s_j r_j \vec{\omega}_j^\text{sf} - s_i r_i \vec{\omega}_i^\text{sf} + s_k r_k \vec{\omega}_k^{fs} - s_j r_j \vec{\omega}_j^\text{sf}   &=  c \vec{x}_{ij} + c \vec{x}_{jk},
\\
s_k r_k \vec{\omega}_k^{fs} - s_i r_i \vec{\omega}_i^\text{sf} &=  c \vec{x}_{ik},
\end{aligned}
\end{equation}
which shows the same relation as Eq.\ (\ref{eqA:srw_ij}) for the non-contacting spheres $i$ and $k$. Therefore, Eq.\ (\ref{eqA:srw_ij}) is valid for any pair of spheres in the assembly.

There are two options of how to extend a 4DOF assembly. One way is to connect two 4DOF assemblies $A$ and $B$. Before they are connected, each of them has an independent parameter $c$ which we call $c_A$ and $c_B$, that describe the slip-free state according to Eq.\ (\ref{eqA:srw_ij}), which holds for any two spheres $i$ and $j$ in the assembly. If we want to form another 4DOF assembly by connecting $A$ and $B$, we need to make sure that the way of connecting enforces $c_A=c_B$ for the slip-free state. This can be done by involving two spheres of each assembly to couple the two parameters $c_A$ and $c_B$. To involve two spheres of each assembly we need to establish at least two contacts.
\begin{figure}[]
\begin{center}
	\includegraphics[width=\columnwidth]{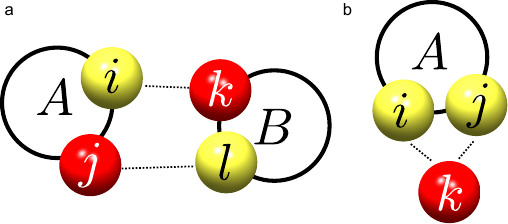}
\end{center}
\caption{
\label{figA:connect} \textbf{Ways of extending 4DOF assemblies.} Every 4DOF assembly can be iteratively constructed using two rules of connecting as minimal requirements. Either connect two 4DOF assemblies $A$ and $B$ by establishing two contacts involving two spheres of each assembly such that the centers of the four spheres $i$, $j$, $k$, and $l$ involved are not coplanar (a). Or integrate a single sphere $k$ in a 4DOF assembly $A$ by establishing two contacts between them such that the centers of involved spheres are not collinear (b).
}
\end{figure}
If we connect sphere $i$ and $j$ from assembly $A$ to sphere $k$ and $l$ from assembly $B$ as shown in Fig.\ \ref{figA:connect}(a), and consider the fact that $A$ and $B$ are 4DOF assemblies with parameters $c_A$ and $c_B$, respectively, we can establish the constraints
\begin{align}
s_j r_j \vec{\omega}_j^\text{sf} - s_i r_i \vec{\omega}_i^\text{sf}  &= c_A \vec{x}_{ij},\\
s_l r_l \vec{\omega}_l^\text{sf} - s_j r_j \vec{\omega}_j^\text{sf} &=  c_{jl} \vec{x}_{jl},\\
s_k r_k \vec{\omega}_k^\text{sf} - s_l r_l \vec{\omega}_l^\text{sf}  &= c_B \vec{x}_{lk},\\
s_i r_i \vec{\omega}_i^\text{sf} - s_k r_k \vec{\omega}_k^\text{sf} &=  c_{ki} \vec{x}_{ki},
\end{align}
which if combined lead to a single constraint
\begin{equation}\label{eqA:constraint_connect4}
c_A \vec{x}_{ij} + c_{jl} \vec{x}_{jl} + c_B \vec{x}_{lk} + c_{ki} \vec{x}_{ki} = \vec{0}.
\end{equation}
Only if one finds three linear independent vectors among $\vec{x}_{ij}$, $\vec{x}_{jl}$, $\vec{x}_{lk}$, and $\vec{x}_{ki}$, i.e., if the centers of the four spheres $i$, $j$, $k$, and $l$ are not coplanar, Eq.\ (\ref{eqA:constraint_connect4}) enforces \mbox{$c_A = c_{jl} = c_B = c_{ki}$}, such that the resulting assembly is a 4DOF assembly. Note that spheres $i$ and $j$ of the bipartite assembly $A$ can be of any color, as long as the final assembly is also bipartite. Another way to extend a 4DOF assembly is to integrate a single sphere $k$ by establishing contacts to two spheres $i$ and $j$ of a assembly $A$, as shown in Fig.\ \ref{figA:connect}(b). The constraints
\begin{align}
s_j r_j \vec{\omega}_j^\text{sf} - s_i r_i \vec{\omega}_i^\text{sf} &= c_A \vec{x}_{ij},\\
s_k r_k \vec{\omega}_k^\text{sf} - s_j r_j \vec{\omega}_j^\text{sf} &=  c_{jl} \vec{x}_{jk},\\
s_i r_i \vec{\omega}_i^\text{sf} - s_k r_k \vec{\omega}_k^\text{sf} &=  c_{ik} \vec{x}_{ki},
\end{align}
can be added to obtain 
\begin{align}\label{eqA:single_sphere_integration}
c_A \vec{x}_{ij} +  c_{jk} \vec{x}_{jk}+ c_{ik} \vec{x}_{ki} = \vec{0}.
\end{align}
Only if the centers of the spheres $i$, $j$, and $k$ are not collinear, Eq.\ (\ref{eqA:single_sphere_integration}) enforces \mbox{$c_A = c_{jk} =c_{ik}$}. In that case, the resulting assembly is a 4DOF assembly. Starting from two single spheres being the most simple 4DOF assembly, one can construct every other 4DOF assembly iteratively using the two presented rules for connecting as minimal requirements. The formation of additional contacts during the process does not change the fact that the resulting assembly is a 4DOF assembly as long as it is bipartite.

\section{Description of the slip-free state for 4DOF assemblies}\label{appendix:DescriptionState}

Since Eq.\ (\ref{eqA:srw_ij}) holds for any pair of spheres $i$ and $j$ in the assembly, we can express the angular velocity of any sphere $j$ as a function of the angular velocity of a reference sphere $i$ as
\begin{equation}
\vec{\omega}_j^\text{sf} = \frac{s_j}{r_j} \left(s_i r_i \vec{\omega}_i^\text{sf} + c \vec{x}_{ij} \right).
\end{equation}
To describe the slip-free state in a general way, we choose the reference sphere $i$ not to be an actual sphere of the assembly, but an imaginary reference sphere with angular velocity $\vec{\omega}_i^\text{sf}=\vec{\Omega}$, $s_i=r_i=1$, and $\vec{x}_{i}=\vec{0}$ to obtain
 \begin{equation}\label{eqA:sync_ang_velocities_baram}
\vec{\omega}_j^\text{sf} = \frac{s_j}{r_j} \left( \vec{\Omega} + c \vec{x}_{j} \right).
\end{equation}

\section{Predicting the slip-free state using $\vec{A}$ and $B$}\label{appendix:PredictionState}

The slip-free state as written in Eq.\ (\ref{eqA:sync_ang_velocities_baram})  can be predicted using the time invariant quantities $\vec{A}$ and $B$ defined as 
 \begin{equation}\label{eqA:BearingMomentum}
 \vec{A} := \sum_i s_i m_i r_i \vec{\omega}_i
 \end{equation}
and
\begin{equation}\label{eqA:HedgehogQuota}
B:= \sum_i s_i m_i r_i\vec{\omega}_i \cdot \vec{x}_{i}.
\end{equation}
We write Eq.\ (\ref{eqA:sync_ang_velocities_baram}) for a sphere $i$, multiply it by $s_i m_i r_i$, and sum over all spheres to end up with $\vec{A}$ on the left hand side and to eliminate $c$ from the right hand side, because \mbox{$ \sum_i c m_i \vec{x}_i = c \vec{M} = \vec{0}$}, since we defined the center of mass $\vec{M}$ of the entire assembly to be the origin of our coordinate system ($\vec{M}=\vec{0}$). We find \mbox{$\vec{\Omega}=\vec{A}/M$}, with the total mass \mbox{$M=\sum_i m_i$}. Second, we multiply (dot product) Eq.\ (\ref{eqA:sync_ang_velocities_baram}) written for sphere $i$ by $s_i m_i r_i \vec{x}_i$ and sum over all spheres to end up with $B$ on the left hand side and eliminate $\vec{\Omega}$ on the right hand side to find \mbox{$c=H/I$}, where \mbox{$I = \sum_i m_i |\vec{x}_i|^2$}. We then formulate the angular velocities of the slip-free state as a function of time invariant quantities only as
\begin{equation}\label{eqA:sync_ang_velocities}
\vec{\omega}_i^\text{sf} = \frac{s_i}{r_i} \left( \frac{\vec{A}}{M} + \frac{B}{I} \vec{x}_{i}\right).
\end{equation}

\section{Different sliding forces lead to different kinetic pathways}\label{appendix:Differentpathways}

\begin{figure}[]
\begin{center}
	\includegraphics[width=\columnwidth]{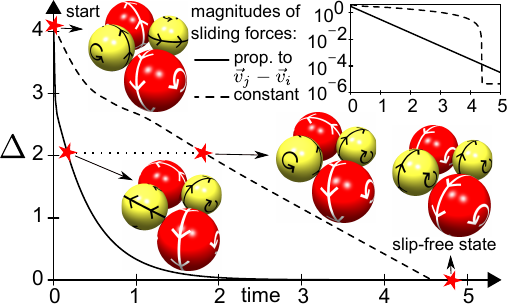}
\end{center}
\caption{
\label{figA:synchronization} \textbf{Different sliding forces merely lead to different kinetic pathways toward the slip-free state.} 4DOF assembly that relaxes from randomly chosen initial angular velocities toward a slip-free state with sliding forces with magnitudes proportional to the difference in tangential velocities (solid) and with constant magnitudes (dashed). \mbox{$\Delta=\sum_i |\vec{\omega}_i - \vec{\omega}_i^\text{sf}|$} measures the deviation from the slip-free state. Inset: Exponential decay of $\Delta$ found for the proportional force. The final slip-free state is independent of the sliding forces in contrast to intermediate states with same $\Delta$. Simulation time in seconds with a time step of $10^{-6}$~s.
}
\end{figure}

The final slip-free state is independent of the type and strength of sliding friction. Different sliding forces only lead to different kinetic pathways toward the final state as illustrated in Fig. \ref{figA:synchronization}, which in general also depend on the geometry of the assembly and the moments of inertia of the spheres.
We show a \textit{noncoplanar-4} loop, a simple 4DOF assembly, relaxing from a random initial configuration toward the slip-free state predicted by Eq.\ (\ref{eqA:sync_ang_velocities}) for two different types of sliding forces. Any sliding force $\vec{F}_{ij}$ acting on sphere $j$ due to contact with sphere $i$ points in the direction opposite to the relative velocity \mbox{$\vec{v}_{ij}=\vec{v}_j-\vec{v}_i$} at the contact. We considered two cases. First, \mbox{$\vec{F}_{ij}= - \sigma \vec{v}_{ij}$}, where the magnitude of each force is proportional to the difference in tangential velocities, with $\sigma=3$ and second, \mbox{$\vec{F}_{ij}= - \sigma \hat{v}_{ij}$}, where the magnitude is constant, with $\sigma=0.05$ and with $\hat{v}_{ij}$ being the unit vector along $\vec{v}_{ij}$.

\section{How blocking a single sphere influences the slip-free state}\label{appendix:BlockingAsphere}

From Eqs.\ (\ref{eqA:BearingMomentum}), (\ref{eqA:HedgehogQuota}), and (\ref{eqA:sync_ang_velocities}), one can derive the new $\vec{A}$ ($\vec{A}_{new}$) and $B$ ($B_{new}$) from the previous (old) values ($\vec{A}_{old}$ and $B_{old}$) when blocking sphere $i$ as described in the following. By applying a perturbation $\Delta \vec{\omega}_i$ to the angular velocity of a sphere $i$ one can impose the changes
\begin{equation}\label{eqA:Delta_i_B}
\Delta_i \vec{A} = s_i m_i r_i \Delta \vec{\omega}_i
\end{equation} 
and
\begin{equation}\label{eqA:Delta_i_H}
\Delta_i B = s_i m_i r_i \Delta \vec{\omega}_i \cdot \vec{x}_i.
\end{equation} 
Remember that the origin of the position vector $\vec{x}_i$ of the center of any sphere $i$ is the center of mass of the entire assembly. Blocking a single sphere permanently and letting the assembly relax to a slip-free state has the same effect on $\vec{A}$ and $B$ as applying a perturbation $\Delta \vec{\omega}_i$ that leads to the sphere $i$ being at rest in the slip-free state, i.e., $\Delta \vec{\omega}_i^\text{sf}=\vec{0}$. To ensure $\Delta \vec{\omega}_i^\text{sf}=\vec{0}$ we find from
\begin{equation}\label{eqA:sync_ang_velocities_freeze}
\vec{\omega}_i^\text{sf} = \frac{s_i}{r_i} \left( \frac{\vec{A}_{new}}{M} + \frac{B_{new}}{I} \vec{x}_{i}\right) = \vec{0},
\end{equation}
that
\begin{equation}\label{eqA:Freeze2}
\vec{A}_{new} = - M B_{new}  \vec{x}_{i} / I.
\end{equation}
From the fact that $\Delta_i \vec{A} \cdot \vec{x}_i = \Delta_i B $ (compare Eqs.\ (\ref{eqA:Delta_i_B}) and (\ref{eqA:Delta_i_H})), we find using $\Delta_i \vec{A} = \vec{A}_{new} - \vec{A}_{old}$ and $\Delta_i B = B_{new} - B_{old}$, that
\begin{equation}\label{eqA:Freeze}
B_{new} = \frac{B_{old} - \vec{A}_{old} \cdot \vec{x}_{i}}{1+M {|\vec{x}_{i}|}^2 /I }.
\end{equation}

\section{Imposing a desired slip-free state}\label{appendix:ImposingState}

We want to impose a desired change $\Delta \vec{A}$ and $\Delta B$ by applying external changes $\Delta \vec{\omega}_i$ and $\Delta \vec{\omega}_j$ to the angular velocity of sphere $i$ and $j$, respectively. Eqs.\ (\ref{eqA:Delta_i_B}) and (\ref{eqA:Delta_i_H}) show the changes $\Delta_i \vec{A}$ and $\Delta_i B$ imposed by the change $\Delta \vec{\omega}_i$. The individually imposed changes in $\vec{A}$ and $B$ need to sum up to the desired change, i.e., \mbox{$\Delta\vec{A}=\Delta_i \vec{A} + \Delta_j \vec{A}$} and \mbox{$\Delta B = \Delta_i B + \Delta_j B$}. We define $\Delta \vec{\omega}_i= \omega_i \Delta \hat{\omega}_i$, where $\hat{\omega}_i$ is the unit vector of the external change $\Delta \vec{\omega}_i$. From Eqs.\ (\ref{eqA:Delta_i_B}) and (\ref{eqA:Delta_i_H}) we find
\begin{equation}\label{eqA:Delta_i_B_plus_Delta_j_B}
\Delta\vec{A}=\Delta_i \vec{A} + \Delta_j \vec{A} =s_i m_i r_i \omega_i \Delta \hat{\omega}_i + s_j m_j r_j  \Delta \vec{\omega}_j
\end{equation}
and
\begin{equation}\label{eqA:Delta_i_H_plus_Delta_j_H}
\Delta B= \Delta_i B + \Delta_j B=s_i m_i r_i \omega_i \Delta \hat{\omega}_i \cdot \vec{x}_i + s_j m_j r_j \Delta \vec{\omega}_j\cdot \vec{x}_j.
\end{equation}
We multiply (dot product) Eq.\ (\ref{eqA:Delta_i_B_plus_Delta_j_B}) with the position vector $\vec{x}_j$ pointing from the center of mass of the assembly to the center of sphere $j$ and subtract Eq.\ (\ref{eqA:Delta_i_H_plus_Delta_j_H}) from it to find
\begin{equation}\label{eqA:omega_i}
\omega_i= \frac{ \Delta\vec{A} \cdot \vec{x}_j - \Delta B}{s_i m_i r_i  \Delta \hat{\omega}_i \cdot \vec{x}_{ij}},
\end{equation}
such that we can choose any $\Delta \hat{\omega}_i$ as long as $\Delta \hat{\omega}_i \cdot \vec{x}_{ij} \neq 0$. From Eq.\ (\ref{eqA:omega_i}) we know $\Delta \vec{\omega}_i$ since $\Delta \vec{\omega}_i= \omega_i \Delta \hat{\omega}_i$ and we obtain using Eq.\ (\ref{eqA:Delta_i_B_plus_Delta_j_B}) that
\begin{equation}\label{eqA:Delta_omega2}
\Delta\vec{\omega}_j=( \Delta\vec{A}-s_i m_i r_i  \Delta \vec{\omega}_i)/(s_j  m_j r_j).
\end{equation}

\section{How to determine the total mass $M$, the center of mass, and the parameter $I$ by accessing only two spheres}\label{appendix:DetermineQuantities}

It is possible to determine all global quantities relevant for the final slip-free state by accessing not more than two spheres of a 4DOF assembly regardless its size. Compared to Eq. (\ref{eqA:sync_ang_velocities}), relevant are the total mass $M$, the center of mass, i.e., the origin of the position vectors $\vec{x}_i$, and the parameter $I$. Starting from any slip-free state, one can separately apply two changes $\Delta_1 \vec{\omega}_{i}$ (first) and $\Delta_2 \vec{\omega}_{i}$ (second) to an accessible sphere $i$. One needs to wait after each change till the slip-free state is reached, and determine the corresponding changes $\Delta_1 \vec{\omega}_i^\text{sf}$ and $\Delta_2 \vec{\omega}_i^\text{sf}$ in angular velocity from the previous to the new slip-free state. Applying a change $\Delta \vec{\omega}_i$ to a sphere $i$ leads to a change $\Delta \vec{\omega}_i^\text{sf}$ of the angular velocity of sphere $i$ between the previous to the new slip-free state. We use
 \begin{equation}
\Delta \vec{\omega}_i^\text{sf} = \frac{s_i}{r_i} \left( \frac{\Delta_i \vec{A}}{M} + \frac{\Delta_i B}{I} \vec{x}_{i}\right)
 \end{equation}
and Eqs.\ (\ref{eqA:Delta_i_B}) and (\ref{eqA:Delta_i_H}) to find
 \begin{equation}\label{eqA:determination_mass_1}
\Delta \vec{\omega}_i^\text{sf} = m_i (\Delta \vec{\omega}_i / M + (\Delta \vec{\omega}_i \cdot \vec{x}_i) \vec{x}_i / I ).
 \end{equation}
 Let us derive two scalar equations from Eq.\ (\ref{eqA:determination_mass_1}) by first squaring it (dot product) to find
 \begin{equation}\label{eqA:determination_mass_2}
\begin{aligned}
|\Delta \vec{\omega}_i^\text{sf}|^2/m_i^2 =  |\Delta \vec{\omega}_i|^2 / M^2 + 2(\Delta \vec{\omega}_i \cdot \vec{x}_i)^2/(MI)+ \\ (\Delta \vec{\omega}_i \cdot \vec{x}_i)^2 |\vec{x}_i|^2 / I^2 
\end{aligned}
\end{equation}
 and second multiplying it (dot product) with $\Delta \vec{\omega}_i$ to find
 \begin{equation}\label{eqA:determination_mass_3}
\Delta \vec{\omega}_i^\text{sf} \cdot \Delta \vec{\omega}_i / m_i = |\Delta \vec{\omega}_i|^2 / M + (\Delta \vec{\omega}_i \cdot \vec{x}_i)^2 / I .
 \end{equation}
Combining the two we can eliminate the term $(\Delta \vec{\omega}_i \cdot \vec{x}_{i})^2$ and   after some rearrangements we find
\begin{equation}\label{eqA:extract_rearranged}
\begin{aligned}
|\Delta \vec{\omega}_i^\text{sf}|^2 / (m_i^2 |\Delta \vec{\omega}_i|^2) = \\ 1/M^2 + (2/M + |\vec{x}_{i}|^2 /I) (\Delta \vec{\omega}_i^\text{sf} \Delta \vec{\omega}_i / (m_i |\Delta \vec{\omega}_i|^2)-1/M).
\end{aligned}
\end{equation}
For a single change $\Delta \vec{\omega}_i$ and its induced change $\Delta \vec{\omega}_i^\text{sf}$ we see in Eq.\ (\ref{eqA:extract_rearranged}) that we have the two unknown quantities $M$ and $|\vec{x}_{i}|^2 /I$. So with two changes $\Delta_1 \vec{\omega}_{i}$ and $\Delta_2 \vec{\omega}_{i}$ and their induced changes $\Delta_1 \vec{\omega}_i^\text{sf}$ and $\Delta_2 \vec{\omega}_i^\text{sf}$ one can find
\begin{gather}\label{eqA:Mcalculation}
\begin{split}
M=(b+\sqrt{b^2-4ac})/ (2a), \\
a=|f_2 g_1 - f_1 g_2|, \ b=|f_2-f_1|, \ c=|g_2-g_1|, \\
f_n = |\Delta_n \vec{\omega}_i^\text{sf}|^2 / (m_i^2 |\Delta_n \vec{\omega}_{i}|^2), \\
g_n = \Delta_n \vec{\omega}_i^\text{sf} \cdot \Delta_n \vec{\omega}_{i} / (m_i |\Delta_n \vec{\omega}_{i}|).
\end{split}
\end{gather}

To locate the center of mass of the assembly we can use Eqs. (\ref{eqA:BearingMomentum}), (\ref{eqA:HedgehogQuota}), and (\ref{eqA:sync_ang_velocities}) to find a vector
\begin{equation}\label{eqA:toward_center_of_mass}
\vec{x}_i' = \Delta \vec{\omega}_i^\text{sf} - m_i \Delta \vec{\omega}_i /M,
\end{equation}
that is parallel to the vector $\vec{x}_i$ pointing from the center of mass to the center of sphere $i$. By applying an additional change $\Delta \vec{\omega}_j$ to a second accessible sphere $j$, one can use Eq.\ (\ref{eqA:toward_center_of_mass}) to obtain a vector $\vec{x}_j'$ parallel to $\vec{x}_j$. The center of mass is the only point in space that can be reached both going along $\vec{x}_i'$ from the center of sphere $i$ and going along $\vec{x}_j'$ from the center of sphere $j$, in the general case where $\vec{x}_i'$ is not parallel to $\vec{x}_j'$. At last, we can solve Eq.\ (\ref{eqA:determination_mass_3}) for $I$ to find
\begin{equation}\label{eqA:Icalculation}
I=\left(\Delta \vec{\omega}_i \cdot \vec{x}_i \right)^2 / \left( \Delta \vec{\omega}_i^\text{sf} \cdot \Delta \vec{\omega}_i / m_i - |\Delta \vec{\omega}_{i}|^2 / M \right).
\end{equation}

\section{How elasticity can lead to torsional forces}\label{appendix:elasticity}

\begin{figure}[]
\begin{center}
	\includegraphics[width=\columnwidth]{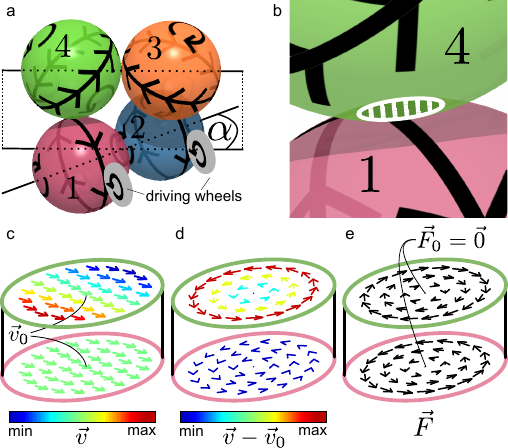}
\end{center}
\caption{
\label{figA:elasticity} \textbf{Elasticity can lead to torsional forces.} (a) \emph{Amplifying mode} of operation: upper spheres (3,4) rotate faster than lower spheres (1,2). (b) the elasticity of spheres leads to a contact area (striped) between spheres instead of point-like contacts assumed in our theory. (c) Velocity profiles of the contact areas between sphere 4 (top) and sphere 1 (bottom). (d) Profile of the velocities after subtracting the velocity at the center of the contact areas. (e) Resulting forces acting on the contact areas lead to a torsional net force on the spheres (length of arrows are proportional to the velocity difference between touching points).
}
\end{figure}

As in the \emph{amplifying mode} of operation shown in Fig.\ \ref{figA:elasticity}(a), elastic spheres can lead to torsional forces due to the formation of finite contact areas between spheres as shown in Fig.\ \ref{figA:elasticity}(b). Let us assume that the velocity of each contact point is proportional to its distance to the rotation axis of the sphere. If not all points of the contact area have the same distance to the rotation axis, the contact area shows an inhomogeneous velocity profile. This is the case for the contact areas between lower and upper spheres in the amplifying mode as shown in Fig.\ \ref{figA:elasticity}(c). The angular velocities are assumed to be the ones predicted by our theory and in this example, we have $\alpha = 36^{\circ}$. To get a better view on the inhomogeneity of the velocity profiles, we plot the difference to the velocity in the center of the contact areas as shown in Fig.\ \ref{figA:elasticity}(d). There one can see that the stronger inhomogeneity of the upper contact area will lead to forces on both contact areas as shown in Fig.\ \ref{figA:elasticity}(e), that sum up to a torsional net force acting on the spheres.

\section{Experimental details}\label{appendix:experimentaldetails}

We used hollow TPU spheres with the trade name \emph{Super High Bounce Jump Ball} with a 10\ cm diameter. The positions of the spheres were fixed using bolt fixing ball transfer units (ball rollers) with 19\ mm carbon steel balls manufactured by \emph{Alwayse Engineering Limited}. For each sphere two to three ball rollers were placed such that with all spheres inserted, all positions stay locked in operation. On the thread of the ball rollers, we placed springs with 37.4\ N spring rate followed by a short brass pipe and a screw nut, such that the ball rollers could be fixed at the pipe, using boss head clamps and rods, and their positions could be slightly adjusted using the screw nut. In the case of too large contact force between the ball roller and a sphere, the spring gives in. Due to the usual application of ball rollers in cargo transfer, they come greased. Since we want to prevent the spheres from slipping on another, we degreased the ball rollers using petroleum. They were one by one dipped into a cup of petroleum such that the ball compartment is covered. Each ball roller was gently patted and rolled on the bottom of the cup for about one minute. Then, before they were dried on air, the petroleum was shaken off and absorbed with household paper. All ball rollers were cleaned in this way four times, where the petroleum was replaced after having washed every ball roller once in it.

To control the rotation of the two lower spheres, we used for each of them a 31.7\ mm large foam rubber tire that we fixed, using a propeller hub, on a gear motor with nine turns per minute and a peak torque of 1.8\ N when operating at 12\ V. The positions of the wheels were fixed such that the contact forces between the wheels and the spheres was large enough to prevent slip during operation and small enough to prevent strong deformation of the spheres.

Every operating setup was caught on video over five minutes. The angular velocities were measured by manual video analysis. Signs were drawn onto the spheres in advance to ease video analysis. We measured the angular velocity as the number of times a sphere turned during the five minutes divided by the exact number of passed seconds. We assumed the rotation axes of the spheres to be fixed over time, which turned out to be not always exactly the case especially for low values of the stacking angle $\alpha$, but a reasonable assumption.

To stack two or three pairs of spheres for a rather simple demonstration, the hollow TPU spheres and the ball rollers fixed in springs worked well. To build assemblies with more spheres, the TPU spheres will turn out to be too soft at some point, and due to compression in different directions depending on the operation mode, they will not suit anymore. For optimal slip-free operation, we suggest spheres to be as light and stiff as possible while having as much grip between spheres as possible.

\end{document}